\documentstyle[preprint,aps]{revtex}

\begin{document}
\draft
\preprint{ }
\title {Current-dependent  exchange-correlation
potential for dynamical linear response theory}
\author{ G. Vignale}
\address {Department of Physics,  University of Missouri, Columbia, 
Missouri 65211}
\author{ Walter Kohn}
\address  {Department of Physics, University of California, Santa
Barbara, California 93106-4030} 
\date{\today} 
\maketitle

\begin{abstract}
The  frequency-dependent exchange-correlation potential, which
appears in the usual Kohn-Sham formulation of a time-dependent linear
response problem, is a strongly nonlocal functional of the {\it
density}, so that a consistent local density approximation generally
does not exist. This  problem  can be avoided  by choosing the {\it
current-density} as the basic variable in a generalized Kohn-Sham
theory. This theory admits a local approximation which, for fixed
frequency, is exact in the limit of slowly varying densities and
perturbing potentials.

\end{abstract}


Gross and Kohn (GK) \cite{GK} have applied the time dependent density
functional theory (TDFT) of Runge and Gross, \cite {Gross}  to the case
of the linear response of interacting electrons in their ground-state to
a time-dependent potential $v_1(\vec r,\omega)e^{-i \omega t}$. The
objective was the 
determination of the linear density response, $n_1(\vec
r,\omega)e^{-i \omega t}$.  They reduced the problem to a set of
self-consistent single particle equations, analogous to the Kohn-Sham
equations for time-independent systems \cite{KS}, with an effective
potential of the form
 \begin {equation} v_1^{eff} (\vec r,\omega) =
v_1(\vec r, \omega) +  \int {n_1(\vec r', \omega) \over \vert \vec r -
\vec r' \vert} d \vec r' + v_{1xc}(\vec r, \omega);
\label {eqK1}
\end {equation}
the exchange-correlation (xc) potential $v_{1xc}(\vec r,\omega)$ is
linear in $n_1(\vec r,\omega)$,
\begin {equation}
v_{1xc}(\vec r,\omega) = \int f_{xc}(\vec r, \vec r';\omega) n_1(\vec
r',\omega) d \vec r',
\label {eqK2}
\end {equation}
and the kernel $f_{xc}(\vec r, \vec r';\omega)$ is a functional of the unperturbed ground state
density $n_0(\vec r)$.

In the spirit of the local density approximation (LDA) for
static and quasi-static problems  \cite {Zangwill} , they then considered the case where
{\it both} $n_0$  and $n_1$ are sufficiently slowly varying functions of
$\vec r$.  As $f_{xc}$ is of short range  for a {\it homogeneous} system,
they proposed the following plausible approximation for systems of
slowly varying $n_0(\vec r)$: 
\begin {equation} f_{xc}(\vec r, \vec
r';\omega)  \sim 
f_{xc}^h(\vert \vec r - \vec r' \vert,\omega;n_0(\vec r)).
\label {eqK3}
\end {equation}
The superscript $h$ refers to a homogeneous electron gas and
the function $f_{xc}^h$ is a property of the homogeneous electron gas
\cite{GK,Conti}. 

However it was noted later by Dobson \cite{Dobson} that
the approximation ~(\ref{eqK3}), when applied to an electron gas in a
static harmonic potential ${1 \over 2}k  r^2$ and subjected to a uniform
electric field, $v_1(\vec r, \omega) = -\vec E \cdot \vec r e^{-i \omega
t}$, violates the so called harmonic potential theorem   (HPT), related
to the generalized Kohn's theorem \cite{Kohn}, according to which the
density follows rigidly the classical motion  of the center of mass: 
$n_1(\vec r,\omega) = \vec \nabla n_0(\vec r) \cdot \vec R_{CM}
(\omega)$.  This raised serious questions about the validity of the
approximation ~(\ref {eqK3}).
Dobson  observed  that one could satisfy the HPT by requiring that the
GK approximation  ~(\ref {eqK3}) be applied in a frame of reference
moving with the local velocity of the electron fluid.
The xc potential obtained by this construction is a  functional of the
{\it current-density} as well as the density \cite{Bunner}.

Further light on the problem with approximation  ~(\ref{eqK3}) was
thrown by Vignale's observation \cite {Vignale} that the 
covariance of the time-dependent Schr\"odinger equation under
transformation to an accelerated frame of reference requires 
the  total force exerted on the system by the  exchange-correlation
and Hartree potentials to vanish, in agreement with the third law of
Newtonian mechanics.  This implies that the exact $f_{xc}$
must satisfy the  sum rule  \cite{Vignale2}  
\begin {equation}
\int f_{xc}(\vec r,\vec r';\omega) \vec \nabla ' n_0(\vec r') d\vec r' =
\vec \nabla v_{0xc} (\vec r), \label {sumrule}
\end {equation}
where $v_{0xc}(\vec r)$ is the static $xc$ potential. This sum rule is
violated by eq. ~(\ref{eqK3}). More generally,
one can deduce \cite{Vignale2} that $f_{xc}(\vec r,\vec r',\omega)$ for a
non-uniform system is of long range in space and  a
nonlocal  functional of the  density distribution. These results 
indicate that,  contrary to more optimistic expectations, a local-density
approximation for time-dependent linear response in general does not
exist as long as one insists
on describing  dynamical exchange-correlation effects in terms of the
density. 

In this paper, we want to demonstrate however that a
local approximation   for the  time-dependent
linear response theory can be constructed in terms of the {\it current
density}.
We consider the linear {\it current} response $\vec j_1(\vec r,
\omega)e^{-i \omega t}$ of interacting electrons
in their ground-state to a time-dependent {\it vector} potential $\vec
a_1(\vec r,\omega) e^{-i\omega t}$.  This problem  includes, as a
special case, the scalar potential problem  studied by GK, because any
scalar potential $v_1(\vec r,\omega)$ can be gauge-transformed to a
longitudinal vector potential $\vec a_1(\vec r,\omega) = \vec \nabla
v_1(\vec r, \omega)/i \omega$,  and the density response can be
calculated from the current response  using the continuity equation
$n_1(\vec r,\omega) = \vec \nabla \cdot \vec j_1(\vec r, \omega)/i
\omega$. As usual, we  express the exact induced current  as
 the response of a non-interacting reference system (the
``Kohn-Sham" system) to an effective vector potential $\vec a_1^{eff}
= \vec a_1 + \vec a_{1H} + \vec a_{1xc}$:
\begin {equation}
j_{1i}(\vec r,\omega) = \int \sum_j {\bf \chi}_{KS,ij} (\vec r,\vec
r',\omega) \cdot \vec a_{1j}^{eff}(\vec r',\omega)d\vec r',
\label {KohnSham}
\end {equation}
where
\begin {equation}
\vec a_{1H} (\vec r,\omega)= {1 \over (i \omega)^2} \int \vec \nabla {1
\over \vert \vec r - \vec r' \vert} (\vec \nabla ' \cdot \vec j_1(\vec
r',\omega )) d\vec r' \label{Hartreepotential}
\end{equation}
is the longitudinal vector potential corresponding
to the dynamic Hartree potential of eq.~(\ref{eqK1}),
 and
\begin {equation}
\vec a_{1xc} (\vec r,\omega) = \int {\bf f}_{xc} (\vec r, \vec
r',\omega) \cdot \vec j_1(\vec r',\omega) d\vec r'
\label{axc}
\end {equation}
is a linear functional of the current.
The $3 \times 3$ tensor kernel ${\bf f}_{xc}$ is defined as
\begin {equation}
{\bf f}_{xc,ij} (\vec r,\vec r',\omega) = 
{\bf \chi}_{KS,ij}^{-1} (\vec r, \vec r',\omega) - 
{\bf \chi}_{ij}^{-1} (\vec r, \vec r',\omega)
-\nabla_{i} {1 \over \vert \vec r - \vec r' \vert} \nabla_{j}',
\label{deffxc}
\end{equation}
where ${\bf \chi}$ and  ${\bf \chi}_{KS}$  are the  current response
tensors \cite{Nozieres}  of the interacting system and the ``Kohn-Sham"
noninteracting system respectively. The latter is  given as
\begin
{equation} {\bf \chi}_{KS,ij} (\vec r, \vec r',\omega) = {n_0(\vec r)
\over m} \delta (\vec r - \vec r') \delta_{ij}+ {1\over
m^2}\sum_{\alpha,\beta} (f_\alpha-f_\beta) {\psi^*_\alpha (\vec r)
\nabla_i \psi_\beta (\vec r) \psi^*_\beta(\vec r') \nabla_j' \psi_\alpha
(\vec r') \over \omega - (\epsilon_\beta - \epsilon_\alpha) +i \eta},
\label {chikohnsham} \end {equation} where $\psi_\alpha (\vec r)$ are 
the solutions of the static Kohn-Sham equation, with eigenvalues
$\epsilon_{\alpha}$. (We
 have put $e=c=1$).  
For a {\it
homogeneous} electron gas of density $\bar n$,  ${\bf f}_{xc}$ is a
function of $\vec r - \vec r'$, which can be Fourier-transformed to 
${\bf f}^h_{xc}(\vec k, \omega)$.  In the limit of small wave vector 
$k$ , which, at fixed frequency $\omega$, means
$k<<k_F,\omega/v_F$  ($k_F =$ Fermi momentum, $v_F =$
Fermi velocity) \cite{smallk},  ${\bf f}^h_{xc}$ has the form  
\begin {equation}
{\bf f}^h_{xc,ij} (\vec k, \omega) = {1\over \omega ^2}
[f^h_{xcL} (\omega,\bar n)  k_i k_j +
f^h_{xcT}
(\omega,\bar n) (k^2 \delta_{ij} -  k_i  k_j)],
\label {fxch}
\end {equation}
where the factor $1/\omega^2$ has been put in evidence so that the
function $f^h_{xcL}(\omega,\bar n)$ coincide with the $k \to 0$ limit
of  $f^h_{xc}(k, \omega,\bar n)$  introduced by $GK$  (see
eq.~(\ref{eqK3})). 
 Both $f^h_{xcL}(\omega,\bar n)$ and $f^h_{xcT}(\omega,\bar n)$ can, in
principle, be   computed from the  current response function of
the homogeneous electron gas (see eq.~(\ref{deffxc})). While  some
aspects of these functions are known   \cite {GK,Footnote0},
further work is needed for a complete evaluation.

From  eq.~(\ref{fxch}) we can obtain the $xc$ potential in real space for
a homogeneous electron gas subjected to a perturbation which is slowly
varying on the scales $k_F^{-1}$ ($\sim$ interelectron
distance) and $v_F/\omega$ ($\sim$ distance travelled by an electron
during a period of the  perturbing field): 
\begin {equation}
\vec a^h_{1xc}(\vec r,\omega) = -{1 \over \omega^2}\{ \vec \nabla
[f^h_{xcL} (\omega,\bar n) \vec \nabla \cdot \vec j_1(\vec r,
\omega)] - \vec \nabla \times [f^h_{xcT} (\omega,\bar n) \vec \nabla
\times \vec j_1(\vec r, \omega)]\}.
\label{naivelda}
\end {equation}

Next, in the spirit of the local density approximation, let  us 
consider a system  in which the unperturbed static density $n_0(\vec r)$
is  slowly varying on the scales of the local $k_F^{-1}$ and
$v_F/\omega$, but not necessarily on the scale of $k^{-1}$, the
wavelength of the perturbing field.

The  simplest case is that of a  periodically modulated electron gas,
where the  unperturbed  density, given by 
\begin {equation}  n_0(\vec r) =
\bar n (1 + 2 \gamma \cos {(\vec q \cdot \vec r)}),  \label{n0} \end
{equation} is not only slowly varying (i.e., $q<<k_F, \omega/v_F$), 
but, also, {\it almost  uniform}, i.e. $\gamma<<1$. 

We shall compute
the exact ${\bf f}_{xc}$ for this system, to first order in $\gamma$.
It is convenient to represent ${\bf f}_{xc}$ in momentum space. 
Translational invariance of the uniform density $\bar n$ gives, to first
order in $\gamma$,  ${\bf f}_{xc}(\vec k,\vec k,\omega) = {\bf
f}^h_{xc}(\vec k,\omega)$ (see eq.~(\ref{fxch})) and ${\bf f}_{xc}(\vec
k  + m \vec q,\vec k,\omega) = 0$ for integers $m$ with $ \vert m \vert
>1$.  It remains to calculate the matrix element ${\bf
f}_{xc}(\vec k +\vec q,\vec k,\omega)$.  

This is facilitated by two exact identities which follow
from the transformation   of the time dependent Schr\"odinger equation
in accelerated reference frames  \cite{Footnote1}. The physical content
of these identities is that the  total force and the  total torque
exerted by the Hartree and $xc$ potentials on the system must vanish, in
accordance with Newton's third law. In the specific instance considered
here,   the identities take the form \begin {equation} 
lim_{k \to 0} {\bf f}_{xc,ij}(\vec k + \vec q ,\vec
k,\omega) = -{\gamma \over \omega^2} [\delta f^h_{xcL} q_i q_j +
f^h_{xcT} (q^2 \delta_{ij} - q_iq_j)],  
\label{fxcq1}
\end {equation}
and
\begin {equation}
lim_{k \to 0} \sum_{j,k} \epsilon_{ljk} {\partial \over \partial
k_k} {\bf f}_{xc,ij}(\vec k + \vec q ,\vec k,\omega) = 
-{\gamma \over  \omega^2}  [\delta f^h_{xcL} - 3f^h_{xcT}]
\sum_k \epsilon_{lki}q_k,
\label{fxcq2}
\end {equation}  
where $\epsilon_{ijk}$ is the  Levi-Civita tensor and we have suppressed
$\omega$ and $\bar n$ in $f^h_{xcL}$ and
$f^h_{xcT}$;  and 
\begin {equation}
\delta f^h_{xcL}(\omega,\bar n) \equiv f^h_{xcL}(\omega,\bar n) -  
f^h_{xcL}(\omega =0,\bar n).
\label {deltafxcl}
\end {equation}
These identities are valid for $k,q << k_F, \omega/v_F$ and to first
order in $\gamma$. The condition on $q$ has been used to replace 
$f^h_{xcL}(q,\omega)$ and
$f^h_{xcT}(q,\omega)$ by their small $q$ limit \cite{smallk}.
A third limiting form of ${\bf f}_{xc,ij}$ is obtained from the usual
Ward identity
\begin {equation}
lim_{q \to 0} {\bf f}_{xc} (\vec k +\vec q,\vec k,\omega) =  \gamma
\bar  n {\partial  {\bf f}^h_{xc} (\vec k,\omega,\bar n) \over \partial
\bar n}. \label{wi}
\end {equation}
Finally, we note the symmetry relation
\begin {equation}
{\bf f}_{xc,ij} (\vec k +\vec q,\vec k,\omega)
={\bf f}_{xc,ji} (-\vec k,-\vec k -\vec q, \omega).
\label {symmetry}
\end {equation}

Equation ~(\ref{symmetry}) requires that, for small $k$ and
$q$, 
${\bf f}_{xc,ij} (\vec k +\vec q,\vec k,\omega)$  be expressible as a
linear combination of $q_iq_j, q^2 \delta_{ij}, (k_i+q_i)k_j,
k_i(k_j+q_j)$, and $\vec k \cdot (\vec k+ \vec q)\delta_{ij}$.  The
equations \ref{fxcq1}, \ref{fxcq2}, and \ref{wi}, completely determine
the coefficients of this combination. Thus, to first order in $\gamma$,
we  obtain

\begin{eqnarray}
{\bf f}_{xc,ij}(\vec k + \vec q, \vec k,\omega)  &=&
-{\gamma \over \omega^2}\{(\delta f^h_{xcL} - f^h_{xcT})q_iq_j
+f^h_{xcT} q^2 \delta_{ij}
-  \bar n {\partial f^h_{xcT} \over \partial \bar n}\vec k \cdot (\vec k
+ \vec q) \delta_{ij}  \nonumber\\ && +
A(\bar n,\omega) (k_i + q_i)  k_j
-B(\bar n, \omega) 
k_i (k_j +  q_j) \},  \label{offdiagonalfxc}  
\end{eqnarray}
where 
$A(\bar n,\omega) \equiv [\bar n (2 \partial f^h_{xcT} /
\partial \bar n  - \partial f^h_{xcL} /
\partial \bar n) + 3f^h_{xcT} - \delta f^h_{xcL}]$, and $B(\bar
n,\omega) \equiv [\bar n \partial f^h_{xcT} / \partial \bar n +
3f^h_{xcT} - \delta f^h_{xcL}]$. The essential point is that
eq.~(\ref{offdiagonalfxc}) is analytic for small $k$ and $q$. It is this
feature that will enable us to construct a local approximation for $\vec
a_{1xc}$ in terms of the current density. (By contrast, the off-diagonal
component  $f_{xc} (\vec k +\vec q, \vec k,\omega)$ in the usual time
dependent density functional theory, has a singularity of the form $\vec
k \cdot \vec q / k^2$ when $k \to 0$ at finite $q$. This small-$k$ 
singularity  reflects the already mentioned long-rangedness of $f_{xc}$
and the consequent nonlocality of the description of $xc$ effects in
terms of the density.)

Let us now return to the more general problem of determining $\vec
a_{1xc}$ for a system whose density is  {\it slowly varying} in the
sense $\vert \vec \nabla n_0(\vec r) \vert / n_0(\vec r) <<k_F(\vec
r)$ and  $\omega /v_F(\vec r) $, but  may have  large global deviations
from uniformity. We assume that $\vec a_{1xc}$ of such a system can be
expanded in a power series of gradients of the local unperturbed
density.  The most general form of $\vec a_{1xc}$ containing up to two
gradient operators is then a linear combination of the following terms:
$\vec j_1, \vec \nabla n_0 \times \vec j_1,     
\vec \nabla \times \vec j_1, \vec \nabla (\vec \nabla \cdot \vec j_1), 
\vec \nabla \times (\vec \nabla \times \vec j_1),
\vec \nabla (\vec \nabla n_0 \cdot \vec j_1),
\vec \nabla \times (\vec \nabla n_0 \times \vec j_1),
\vec \nabla n_0 (\vec \nabla \cdot \vec j_1),
\vec \nabla n_0 \times (\vec \nabla \times \vec j_1),
(\vec \nabla n_0 \cdot \vec \nabla) \vec j_1,
\vec \nabla n_0 (\vec \nabla n_0 \cdot \vec j_1),
\vec \nabla n_0 \times (\vec \nabla n_0 \times \vec j_1)$
with coefficients that are functions of $n_0(\vec r)$ and $\omega$.
This gradient expansion is applicable, in particular, to the
periodically modulated electron gas of eq.~(\ref{n0}), in which case it
must yield the  same  perturbative (in $\gamma$) result  that one
obtains from the combination of eqs.~\ref{axc}, \ref {fxch}, and
\ref{offdiagonalfxc}.  Thus, by requiring agreement between the gradient
and the perturbative expansions,  we are able to  determine the
coefficients of all the terms appearing in the gradient expansion,
except the last two, which are of order $\vert \vec \nabla n_0 \vert^2$,
or $\gamma^2$ in the periodically modulated electron gas. We note that
the part of $\vec a_{1xc}$ which is  determined by this procedure
satisfies the HPT up to corrections of order $\vert \vec \nabla n_0
\vert^2$. To fix the coefficients of the second order terms, an
additional condition  must be  imposed. We  require that, if the system
is subjected to an external field, which causes  it to  translate {\it
rigidly} as a whole, i.e.,  $j_1(\vec r,\omega) = n_0(\vec r) \vec v
(\omega)$, 
then $\vec a_{1xc}$ must reduce to $-\vec \nabla [f_{xcL}(\omega=0)\vec
\nabla \cdot j_1(\vec r ,\omega)]/\omega^2$,  a
rigid displacement of the {\it static} exchange-correlation potential
\cite {Vignale}. In the special case of 
electrons confined by a harmonic potential and subjected to a uniform
electric field, this condition guarantees that the HPT is satisfied. 

 The final result of our analysis, obtained after
lengthy but elementary manipulations, is the complete form of the local
approximation for     $\vec a_{1xc}(\vec r,\omega)$  up to second order
in the gradient expansion:   \begin {eqnarray} \vec
a_{1xc}(\vec r, \omega) &=& -{1 \over\omega^2} \{ \vec \nabla [f^h_{xcL}
\vec \nabla \cdot \vec j_1(\vec r, \omega) -  \delta f^h_{xcL} \vec
\nabla n_0(\vec r) \cdot {\vec j_1(\vec r, \omega) \over n_0(\vec r)}]
 - \vec \nabla \times [f^h_{xcT}n_0 \vec
\nabla \times {\vec j_1 (\vec r,\omega) \over n_0(\vec r)}]
\nonumber\\ &&
+ \delta f^h_{xcL} \vec \nabla  n_0(\vec r) \vec \nabla
\cdot {\vec j_1(\vec r, \omega) \over n_0(\vec r)}+f^h_{xcT} [(\vec
\nabla n_0(\vec r)  \cdot \vec \nabla) {\vec j_1(\vec r, \omega) \over
n_0(\vec r)}
-4 \vec \nabla  n_0(\vec r)\vec \nabla
\cdot {\vec j_1(\vec r, \omega) \over n_0(\vec r)}
\nonumber\\&&
+ 3 \sum_j \nabla_j n_0(\vec r) \vec \nabla
{j_{1j}(\vec r, \omega) \over n_0(\vec r)}]
+2 n_0(\vec r) [\sum_j
\nabla_j f^h_{xcT}  \vec \nabla { j_{1j}(\vec r, \omega) \over n_0(\vec
r)} -\vec \nabla f^h_{xcT}\vec \nabla
\cdot {\vec j_1(\vec r, \omega) \over n_0(\vec r)}] \},
\label{finalresult} 
\end {eqnarray}
where $f^h_{xcL}$ and $f^h_{xcT}$  are functions of the local
density $n_0(\vec r)$ and the frequency $\omega$.

Our result simplifies considerably in two limiting cases.
(1) The density is slowly varying on the scale  set by the wavelength
of the perturbation. In this regime  we can neglect the terms containing
$\vec \nabla n_0$, and we recover the former results of GK \cite{Gross}
for the response to a scalar potential, and of Ng \cite{Ng} for the
response to a general vector potential. This result  can also be simply
obtained from eq.~(\ref{naivelda}) by  the substitution $\bar n
\to n_0(\vec r)$. (2) The velocity field $j_1/n_0$ is
constant in space.  This is the case when the static external potential
is parabolic and the perturbing electric field is uniform, the  regime
of Kohn's theorem \cite{Kohn} and the HPT \cite{Dobson}. Then all
 derivatives of $j_1/n_0$ vanish, and eq.~(\ref{finalresult}) reduces to 
\begin {eqnarray} \vec a_{1xc}(\vec r, \omega) &=& -{1 \over\omega^2} \{
\vec \nabla [f^h_{xcL} \vec \nabla \cdot \vec j_1(\vec r, \omega) - 
\delta f^h_{xcL} \vec \nabla n_0(\vec r) \cdot {\vec j_1(\vec r, \omega)
\over n_0(\vec r)}].    \label {Dobson} \end {eqnarray} By a gauge 
tranformation, this
longitudinal  vector potential  can be transformed to
the scalar $xc$ potential proposed by Dobson \cite{Dobson} to satisfy
the HPT. However, for the general case of slowly varying $n_0(\vec r)$
and $\vec a_1(\vec r, \omega)$, the full expression
(\ref{finalresult}) is required. 

 In conclusion, our analysis uniquely specifies  a local
current density functional theory of 
the linear  current (and density) response, which becomes exact in
the limit of slowly varying unperturbed densities and perturbing
potentials. The scale on which the variations  must be slow is set by the
smaller of the wave vectors $k_F$ and $\omega/v_F$. Therefore, this
theory is applicable to the study of high frequency phenomena, such as
electromagnetic  absorption,  for which the adiabatic approximation
\cite{Zangwill} is in general not justified. A complete local current
density response theory, for spatially slowly varying unperturbed
density and perturbing field, and {\it all} frequencies ($0<\omega <
\infty$) remains to be developed.

 We acknowledge support from NSF grants No. DMR-9403908 and DMR-9308011.
One of us (GV) gratefully acknowledges the 
hospitality of the  Institute of Theoretical Physics and of the Physics
Department at UCSB, where  this work was initiated.  We thank A. K.
Rajagopal for  several stimulating discussions.


\begin{references}
\bibitem {smallk} The response functions of the homogeneous electron
gas change behavior as $k$ crosses over from
$k < \omega/v_F$ to $k > \omega/v_F$. Our ``small $k$" limit refers to
the  first regime.
\bibitem{Footnote0} We have
obtained a few exact results about the function  $f^h_{xcT}(\omega, n)$
similar to the ones already known for $f^h_{xcL}(\omega, n)$. To first
order in the Coulomb interaction, $f^h_{xcT}(\omega, n)$ is independent
of frequency and equal to $3 \pi e^2 / 10 k_F^2$. The limit of
$f^h_{xcT}$  for $\omega \to \infty$ is given by ${2 \over 5} n^{2/3}{d
\over dn} {\epsilon_{xc}(n)\over n^{2/3}} + 2 n^{1/3} {d \over dn}
{\epsilon_{xc}(n)\over n^{1/3}}$ where $\epsilon_{xc}(n)$ is the
exchange-correlation energy per particle of the uniform electron gas of
density $n$.
\bibitem{Footnote1} A detailed derivation and discussion of these
``acceleration identities" will be published elsewhere.
\end{references}
\end{document}